# AC loss and coupling currents in YBCO coated conductors with varying number of filaments

Eduard Demencik, Michal Vojenciak, Anna Kario, Rainer Nast, Alexandra Jung, Wilfried Goldacker, Francesco Grilli

Corresponding author: francesco.grilli@kit.edu

**Abstract.** Striation of HTS coated conductors as a way to reduce their magnetization AC losses has been the subject of intense worldwide research in the past years by several groups. While the principle of this approach is well understood, its practical application on commercial material to be used in power application is still far to be implemented due to manufacturing and technological constraints. Recent advances in tape quality and striation technology are now enabling systematic investigations of the influence of the number of filaments on AC loss reduction with a consistency that was not available in the past.

In the present work we demonstrate the technological feasibility of importantly reducing the magnetization losses of commercially available coated conductor by striating them into a high number of filaments (up to 120). The loss reduction exceeds one order of magnitude and does not come at the expense of current-carrying capability: samples with 10 and 20 filaments are unaffected by the striation process, while samples with 80 and 120 filaments still retain 80 and 70% of the current-carrying potential, respectively. We also investigate the transverse resistivity between the filaments in order to understand the paths followed by the coupling currents: we found that the coupling current prevalently flows in the metallic substrate, rather than in and out of the filaments. Finally, we use oxidation as a method to reduce the coupling currents and the corresponding losses.

The contribution of this work is three-fold: 1) It describes the know-how to produce a large number of high quality striations in commercially available coated conductor, greatly reducing their losses without extensively degrading their transport properties; 2) It provides a comprehensive characterization of said samples (e.g. measurements in a wide frequency range, transverse resistance profiles, influence of oxidation on DC and AC behavior of the sample); 3) It provides new insight on the patterns of the coupling currents.

## 1. Introduction

Rare-earth based coated conductors are currently regarded as the most promising high-temperature superconductor (HTS) tapes because of their high critical current density, which is maintained also at very high field and with a low degree of anisotropy thanks to targeted pinning techniques [1]. Coated conductors typically consist of a very thin (about 1 μm) superconducting layer deposited on sub-micron buffer layers grown on a flexible metallic substrate. These tapes are currently being manufactured in long lengths, and critical currents exceeding 300 A in self-field at 77 K are now routinely obtained on 12 mm wide samples of several hundred meters long. However, for a wide class of applications involving windings (such as transformers, motors and generators), the AC loss level of these tapes is too high, essentially because of the negative effect of perpendicular field on flat tapes with high aspect ratio such as coated conductors.

Magnetization losses of thin superconducting strips increase with the square of the strip's width [2], [3], so an obvious way of reducing the losses is by subdividing the superconducting film into stripes (filaments). In order for the process to be effective, however, the filaments need to be twisted or transposed, because of the currents coupling the filaments, either at the current leads or along the length of the tape. The CORC cable design has recently shown promising results in this respect [4].

In the past few years, several groups have investigated the effectiveness of creating filaments by laser ablation for reducing AC losses. This has been done with various techniques: laser scribing [5]–[10], chemical etching [11], [12], mechanical scratching of the buffer layers [13], lift-off process with photoresist mask technique [14], dry etching [15], inkjet printing [16] or the combination of laser scribing and chemical etching [12]. Most of these techniques are developed for short samples only, with lengths between 10 cm and 1 m. The exception is [17], where the authors developed a reel-to-reel striation process for 500 m of tape up to 5 filaments.

Reviewing all the published works is beyond the scope of this work and would require a dedicated publication. Here below we summarize what we believe are the most relevant works in terms of use of this technique in applications-ready coated conductors.

Amemiya *et al.* [18] obtained a loss reduction of a factor about 10 on 20-filaments tapes at low frequency (11.3 Hz). Increasing the frequency, the contribution of the coupling loss becomes important, but at a frequency as high as 171 Hz the AC loss reduction by striation is still apparent. They also estimate the coupling length to be much longer than the analyzed samples (100 mm long), which supports their observation of a detectable AC loss reduction even at high frequency. Majoros *et al.* prepared striated samples with 20 filaments, with a loss reduction of roughly one order of magnitude [6]. They also manufactured samples with bridges between the filaments to favor current sharing between the filaments: the AC losses of samples with multiple bridges were higher than those of the samples with non-bridged filaments due to a significant filament coupling. However, the losses were still about a factor of 2 lower than those of non-striated sample.

Levin *et al.* prepared samples with varying number of filaments (up to 172) on 12 mm wide tape with copper stabilization, but without a very effective reduction of the losses, probably due to filament recoupling caused by the copper [19]. Kesgin *et al.* obtained a loss reduction of a factor 15 on samples with 12 filaments prepared by selective electroplating [20], [21]: after laser ablation, the samples were heat-treated in oxygen atmosphere to oxidize metals in the groove and reduce coupling losses; the procedure was followed by copper stabilizer electrodepositing.

In the present work we demonstrate that it is possible to obtain samples with narrow grooves and large number of filaments (up to 120) that exhibit a substantial AC loss reduction (up to a factor 40) without importantly losing their current-carrying capability; we also investigate the transverse resistivity between the filaments in order to understand where the coupling currents flow. Finally, we use oxidation as a method to reduce the coupling currents and the corresponding losses.

2. **Sample preparation**

The samples were prepared from 12 mm wide Superpower tape, which consists of multilayer oxide buffer structure, ion beam assisted deposition of MgO, *RE*BCO layer, Ag coating 2 μm, 50 microns Hastelloy substrate, without copper stabilization. We used an infrared Nd:YAG-laser with a wavelength of 1030 nm and pulsed with a frequency of 400 kHz. The pulse duration is less than 10 ps. The maximum power of the laser is 25 W, with maximum pulse energy of 125 μJ. For the structuring of the samples we used only 20% of the power (5 W) with 10 repetition cycles for each single groove. For the scribing process a combination of a scanner optic with a speed of 90 m/min and a moving table with 22.5 m/min was used. The spot diameter is 18 μm. The resolution of the repeatable track is 0.5 μm. More details on the scribing process can be found in [22].

Samples with 10, 20, 40, 60, 80 and 120 filaments on 20 cm long tapes were prepared (Figure 1). The last 2.5 cm at both ends of each tape were not striated, in order to have good contacts for current injection. After transport critical current measurement, the ends were cut by laser and used for scanning electron microscopy (SEM) examination (Figure 1). The remaining part was used for AC loss measurements. The width of the laser groove was measured for all prepared samples and found to be equal to 18-21 μm. All the measurements presented in this paper were performed at 77 K.

Another set of samples was heat-treated in $O_2$ atmosphere at 500 °C for one hour for successive DC and AC characterization. The purpose of this operation was to increase the resistance between filaments to limit their coupling, as it is explained later in the text. The temperature and the speed of this oxygenation process were chosen on the basis of a previous investigation on face-to-face silver diffusion joints [23].

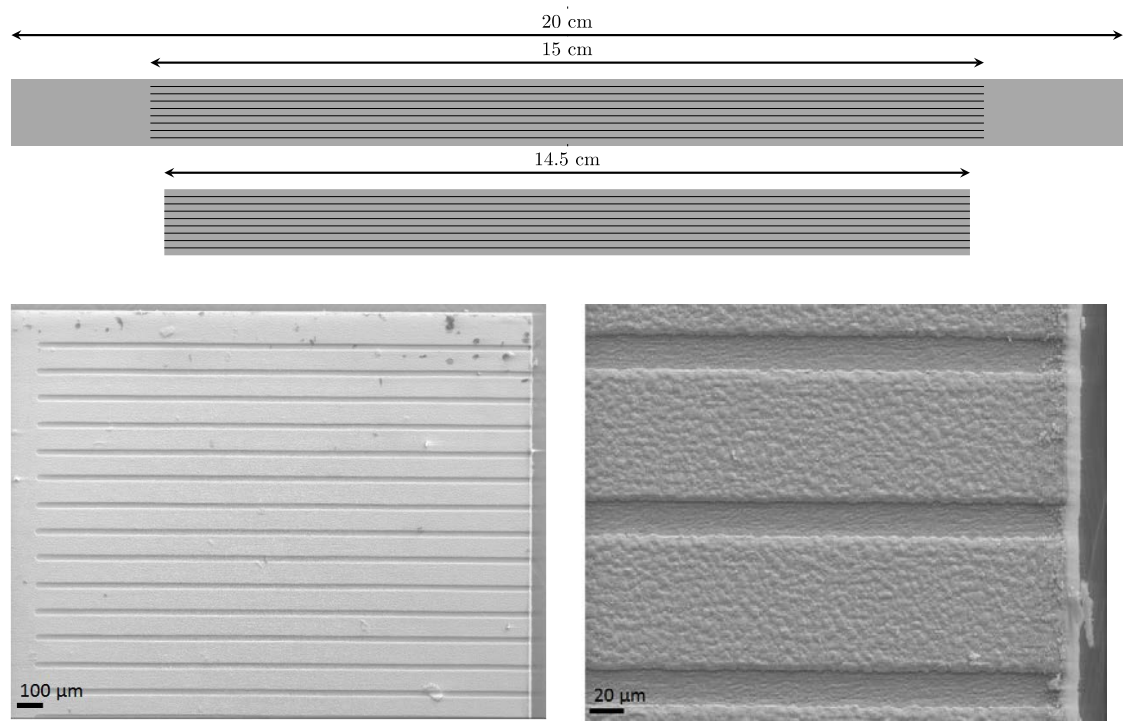

**Figure 1.** Top: schematic drawing of the original 20 cm-long tape (with the 15 cm-long filaments) and the 14.5 cm long sample prepared for AC loss measurements. Bottom: SEM images of the 120-filament tape, with a magnified view on the right.

### 3. DC and AC characterization of the samples

First, we measured the critical current and the $n$ factor of the voltage-current characteristics on samples with different number of filaments (Figure 2). The critical current of the sample with 120 filaments is 187.5 A, which represents 55% of the critical current of the original tape (342 A) and 68% of the value one would expect from the mere loss of superconducting material (274 A). In the sample with 120 filaments the grooves represent around 20% of the sample width. The decrease of the current-carrying capability is less pronounced for samples with fewer filaments: the samples with 10 and 20 filaments are essentially unaffected by the striation process and retain 97% of their current carrying potential (which also indirectly confirms the uniformity of the critical current of the different filaments); the sample with 80 filaments still retains 79% of its potential. These findings encouragingly indicate that the striation process does not dramatically decrease the current carrying capability of samples with several tens of filaments. The striation process has however an impact on the shape of the current-voltage characteristics: the index factor $n$ decreases with increasing number of filaments, which can be put in relation with the increased influence of local defects as the filaments become narrower [24]. The critical current of the samples was not affected by the oxidation process – see open symbols in Figure 2.

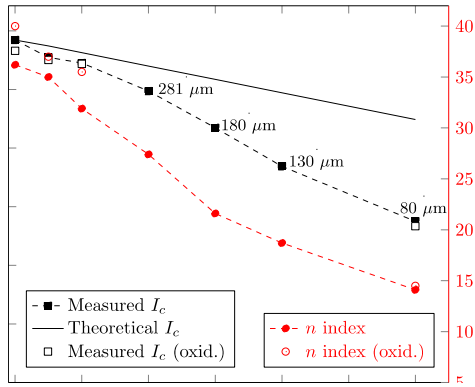

**Figure 2.** Measured critical current (black squares) and n index (red circles) as a function of the number of filaments (whose width is also indicated for some cases). Open symbols represent the same quantities after oxidation. The continuous line indicates the critical current one would expect based on the mere loss of superconducting material.

We subsequently measured the magnetization AC losses of the different samples with the calibration-free method [25]. For this purpose the tapes were cut as shown in Figure 1, so that the filaments run along the whole length of the tape. Figure 3 shows the continuous reduction of the losses with increasing number of filaments for a frequency of 40 Hz. The loss of the original sample closely follows the analytical predictions for a thin strip with $I_c$=340 A. The striated samples have lower losses. If the filaments were perfectly decoupled and the hysteresis loss the only source of dissipation, the loss in the high-field region should reduce proportionally to the number of filament; the experimental data show that this occurs only for the 10-filament tape and in general the loss reduction is more moderate (reaching a factor 40 for the 120-filament tape), which can be ascribed to some degree of coupling between the filaments. It is also interesting to note that at low fields the striated samples have higher losses than the original sample. This is because at low fields the volume of superconductor penetrated by the field is higher in uncoupled filaments than in a non-striated tape [15], [26]. The observed reduction of the slope of the loss curve was confirmed by finite-element method (FEM) simulations (Figure 4). The latter were performed with the 2-D *H*-formulation of Maxwell equations, using a power-law resistivity for the superconductor and imposing complete uncoupling between the filaments by means of appropriate current constraints – see [12] for details on how to impose coupling/uncoupling.

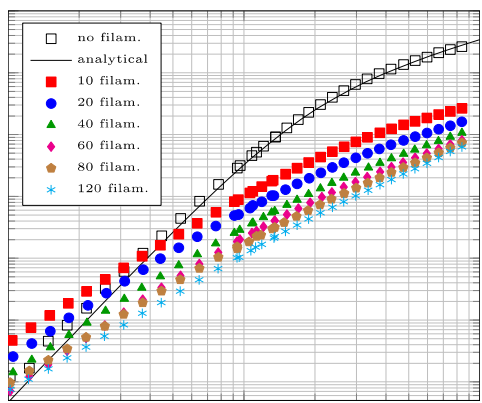

**Figure 3. Magnetization AC losses (at 40 Hz) of samples with different number of filaments.**

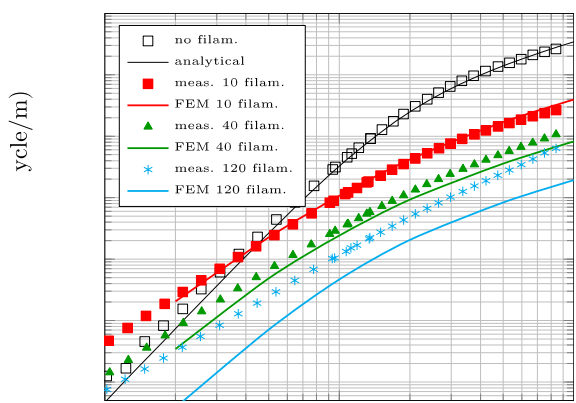

**Figure 4. Comparison with losses computed by FEM simulations assuming perfect uncoupling between the filaments. For sake of readability, only the data for 10, 40 and 120 filaments are shown.**

The agreement between simulations (which assume complete filament uncoupling) and experiments is very good for the 10-filament sample, but it gradually worsens as the number of filaments increases: for the 40-filament sample, the computed losses at 80 mT are 30-40% lower than the measured ones; for the 120-filament sample the discrepancy reaches 70-80%. This discrepancy between measurements and simulations based on perfect uncoupling is another indication that some filament coupling occurs, and that the coupling effect is more important as the number of filaments increases.

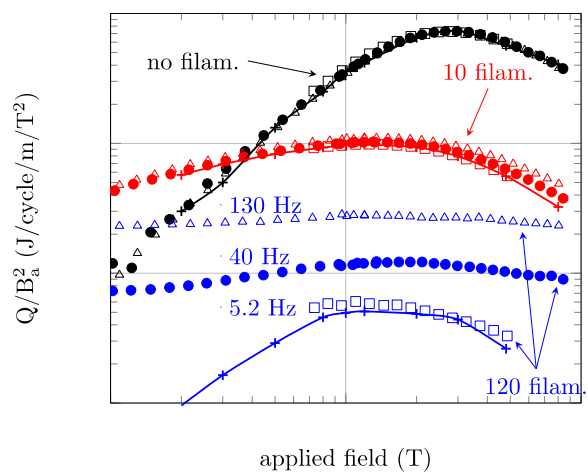

**Figure 5. AC losses normalized by the square of the applied field amplitude for different frequencies and different samples. Different frequencies are represented by different symbols: empty squares (5.2 Hz), full circles (40 Hz), and empty triangles (130 Hz). Different samples are represented by different colors: black (no filaments), red (10 filaments) and blue (120 filaments). The continuous curves represent the zero-frequency limit, for which no coupling exists.**

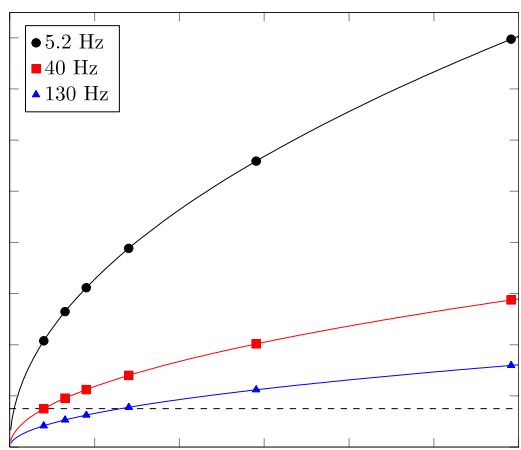

**Figure 6. Critical length as a function of the filament half width according to Wilson's formula for a field of 80 mT and a frequency of 5.2 Hz (black), 40 Hz (red), 130 Hz (blue). The area above each line represents the fully coupled situation. The points represent the critical length for the half width of the samples, which ranges from 40 µm (120 filaments) to 591 µm (10 filaments). The dashed horizontal line represents the length of the sample, which lies almost always below the critical lines (meaning that fully coupling has not been reached). Only for very narrow filaments it is above the blue triangles line, which corresponds to the fully coupling observed for 130 Hz.**

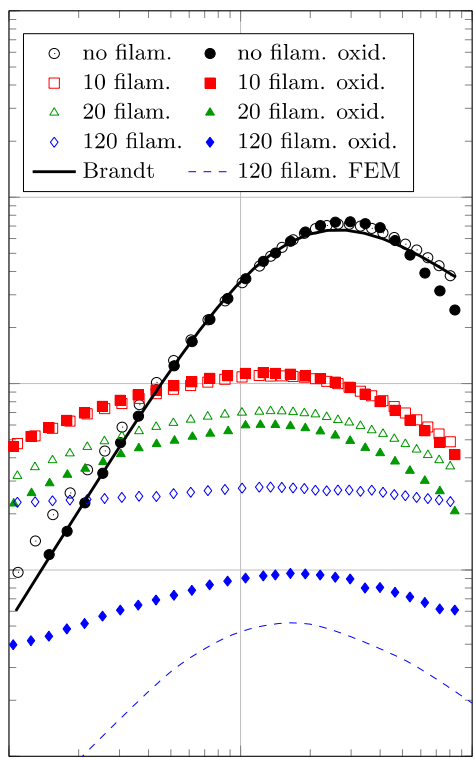

Figure 7. Normalized AC losses at 130 Hz of the original non-striated sample and of samples with 10, 20 and 120 striations, before (empty symbols) and after (filled symbols) oxidation. The black continuous curve represents the losses of the original tape calculated with Brandt's model. The dashed curve in the lower part of the graph represents the limit of completely uncoupled filaments, calculated with FEM calculations. The oxidation helps significantly reduce the degree of coupling; the effect is most obvious in the 120-filament sample.

Filament coupling is also expected to occur more easily at higher frequencies. This is confirmed by the experimental data of Figure 5, which shows the losses normalized by the square of the applied field amplitude for different frequencies and different samples. For the original tape the loss curves at different frequencies nicely overlap; for the 10-filament tape, they start to separate; for the 120-filament tape, they are completely separated. In this latter case, the high-frequency (130 Hz) curve is almost flat, i.e. independent of the field amplitude: since the coupling losses depend on the square of the magnetic field amplitude (see for example formula (5) in [5]), this means that at 130 Hz the filaments are almost fully coupled.

In the figure, the continuous curves represent the zero-frequency limit, calculated by extrapolating the experimental data at zero frequency, for which no coupling exists. For the 120-filament tape the losses measured at 5.2 Hz are very close to this limit. These results tell us that for the 120-filament sample one can pass from full filament uncoupling to full coupling by varying the frequency range between 5 and 130 Hz. For samples with fewer filaments, this frequency range has to be extended.

The observed onset of coupling is also supported by looking at the coupling critical length between two superconductors derived by Wilson [28], although the formula was developed for superconducting slabs in parallel field:

$$l_c = 4\sqrt{\frac{a\,\rho J_c}{dB/dt}} \qquad (1)$$

where $a$ is the filament's half-width, $\rho$ is the resistivity of the inter-filament material, and $dB/dt$ is the variation rate of the field. This latter can be in this case estimated as $2\pi f B_m$, where $B_m$ is the field amplitude. The resistivity appearing in the formula is for a buffer material between the superconducting filaments, whereas in our case the material extends beneath them. This means that in the formula one should consider an effective resistivity of a filament-segregating material positioned between the filaments and with their same thickness, as explained in [18]. Given the thickness of 1 and 50 μm for the superconductor and the substrate, respectively, one should consider a resitivity 50 times lower than that of Hastelloy. As shown in Figure 6, the length of the sample (15 cm) is always lower than the critical length, except for the case of 130 Hz and small filament size: in that case filaments are fully coupled, which further confirms the findings of Figure 5.

The magnetization AC losses were measured also after oxidizing the samples. Figure 7 shows a comparison of the magnetization losses of different samples before and after oxidation. The losses of the original tape and of the 10-filament samples are essentially unaffected by oxidation, whereas as the number of filament increases the positive effect of oxidation can be seen, particularly on the 120-filament sample: the shape of the loss function is no longer flat as before oxidation (when coupling losses were dominant), but becomes slightly peaked, which indicates an increasing importance of the hysteresis losses in the superconductor material. One can also observe that the curve of the oxidized 120-filament sample tends toward the theoretical limit (computed by numerical calculations) of uncoupled filaments.

These results demonstrate the effectiveness of the oxidation process to reduce the coupling losses, which can be put in relation with an increase of the transverse resistance, as explained in the next section.

The fact that, without oxidation, the measured losses are mostly due to the filament coupling can be understood by separating the different loss contributions with the method proposed by Levin *et al.* in [19], which we summarize here for the reader's convenience.

The total dissipated power [e.g. W/m] in a striated sample consists of three contributions, hysteresis, coupling and eddy current losses:

$$P_{tot} = P_h + P_c + P_{eddy} \qquad (2)$$

The power loss per unit of length, sweep rate and critical current can be written as a function of the sweep rate as

$$P/(I_c \cdot B \cdot f) = \lambda_1 + \lambda_2 \cdot B \cdot f \qquad (3)$$

where $\lambda_1$ and $\lambda_2$ are measure of the hysteresis and coupling loss, respectively. Figure 8 shows this type of data analysis for the 120-filament sample at 130 Hz, before and after oxidation. The critical current of the 120-filament sample is $I_c$ = 187.5 A. To determine the parameters $\lambda_1$ and $\lambda_2$ in equation (3) we used a linear fit of the plotted data at high sweep rates.

The parameter values are as follows:

- Sample before oxidation: $\lambda_1$ = 0.083 mm, $\lambda_2$ =0.089 mm.s.T$^{-1}$
- Sample after oxidation: $\lambda'_1$ = 0.068 mm, $\lambda'_2$ =0.019 mm.s.T$^{-1}$.

The parameter $\lambda_1$ represents the hysteresis loss and, in the ideal case of striations with perfect filament separation and no coupling, it equals the width of filament [19]. The parameters $\lambda_1$ and $\lambda'_1$ are equal to 83 and 68 µm, respectively, which are values reasonably close to the measured filament width (80 µm). This is an indication of the positive effect of oxidation for obtaining uncoupled filaments.

The parameter $\lambda_2$ measures both eddy and coupling loss contributions. In order to separate them, one can calculate the eddy current loss in substrate as (see equation (6) in [19]):

$$\lambda_{2,eddy} = \pi^2 \cdot d \cdot W^3 / (6 \cdot \rho_{HS} \cdot I_c) \qquad (4)$$

using the following values: $\rho_{HS}$ = 1.24 .10$^{-6}$ Ω.m (from [29]), $d$=50 µm, $W$=12 mm. This yields a value $\lambda_{2, eddy}$ = 6.11 .10$^{-4}$ mm.s.T$^{-1}$ that is much smaller than the total loss values of $\lambda_2$=0.083 and $\lambda'_2$=0.017 mm.s.T$^{-1}$. This shows that the main contribution to the non-hysteretic loss is coupling loss originating from the resistive connections between the filaments, rather than eddy loss in substrate itself (see also [19]).

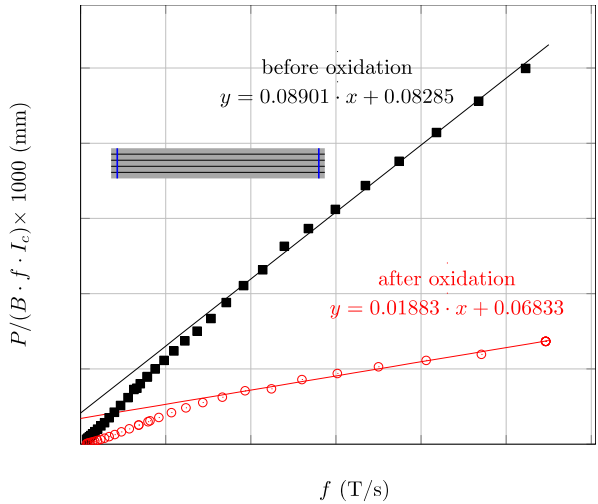

**Figure 8.** Power loss per unit of length, sweep rate and critical current as a function of the sweep rate (see Eq. (2)) for the 120-filament tape at 130 Hz. Shown are the curves before and after heat treatment in $O_2$ at 500°C/1h. The linear fits (taken on the data points with $Bf$ > 3 T/s) are used to determine the parameters $\lambda_1$ and $\lambda_2$ of Eq. (2), which correspond to the hysteretic and non-hysteretic loss contributions, respectively.

## 4. Transverse resistance

Although the onset of coupling is well visible from the analysis of AC loss characteristics carried out in the previous section, less clear is to understand how the coupling currents flow inside the coated conductor. In order to get an insight on this, we measured the transverse resistance at 77 K. For this purpose, in order to avoid the presence of possible reconnecting paths at the end of the samples due to melted parts caused by the laser scribing, we made an additional striation perpendicular to the previous ones very near (and parallel to) the end of the tape (as indicated by the blue lines in Figure 8).

We measured the transverse resistance as follows: we forced a small DC current (up to 0.1 A) to enter the sample in the first filament and to leave it from the last filament. We measured the voltage between the first filament and the other filaments. From the generally linear I-V curves we estimated the resistances. The results for samples with 10 and 20 filaments are shown in Figure 9. The *x*-coordinate represents the position of the centers of the filament along the width of the sample.

When considering the possible paths of the current, the simplest assumptions we can make are that (a) the current flows in the substrate along the whole width of the sample or (b) the current flows in an out of each filament and through the substrate only in the groove [30]. At a given distance $x$ from one of the tape's edges, the resistance offered by the Hastelloy is

$$R_{HAST}(x) = \rho_{HAST} \frac{x}{d \cdot l} = 1.24 \cdot 10^{-6} \frac{x}{5 \cdot 10^{-5} 15 \cdot 10^{-2}} = x \cdot 0.165 \; [\Omega/m], \quad (5)$$

where $d$ is the thickness of the Hastelloy layer (50 μm) and $l$ is the tape length (15 cm), so that their product represents the transverse cross-section of the Hastelloy. The resistance offered by the Hastelloy along the whole width is therefore 12 mm * 0.165 Ohm/m = 1.98 mΩ. The measured value of the transverse resistance is about 3.5 mΩ both for the 10- and 20-filament sample. This value is not too distant from the calculated resistance of 1.98 mΩ, which seems to indicate that the current substantially flows in the Hastelloy, as according to option (a).

In fact, if the current flowed in and out of each filament, we should first of all observe a linear increase of the resistance along the whole width of the sample (as suggested in [31]), which is not observed in the experiments (see Figure 9). In addition, assuming a groove with width 20 μm, this resistance would be equal to 29.7 and 62.7 μΩ for the 10- and 20-filament sample, respectively, which is two orders of magnitude lower than the measured one.

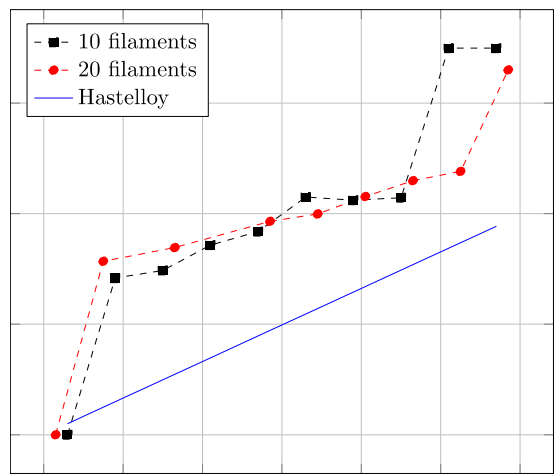

**Figure 9. Measured transverse resistance as a function of the lateral position for the 10 and 20 filament samples. The dashed line represents the expected resistance if only Hastelloy was considered.**

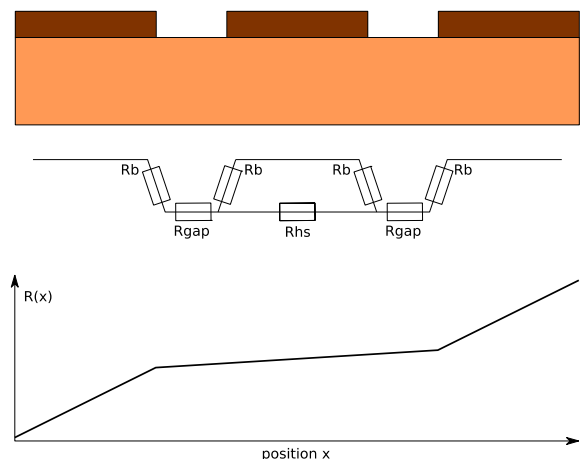

**Figure 10. Network circuit of the various resistive interfaces determining the current paths in a striated coated conductor (for sake of clarity, only three filaments are shown).**

As displayed in Figure 9, both examined samples show a similar pattern: the lateral edges show a sharp increase of the resistance, whereas the central part of the sample is characterized by a moderate increase of the resistance with the position. The slope of the central part is of 116 mΩ/m and of 93 mΩ/m for the 10- and 20-filament samples, respectively. These values are not too far from the value for Hastelloy calculated above (165 mΩ/m).

The analysis carried out above supports the following: the current is forced to enter the substrate through an interface/barrier resistance between the first filament and the substrate. Then it is free to either flow in and out of the each filament or stay in substrate. Similarly, the current is forced to flow through an interface/barrier resistance into the last filament. This is schematically illustrated by the network circuit of Figure 9 (see also [18]): A sharp step-like increase of resistance is expected if the interface resistance is much higher than the resistance of the Hastelloy substrate: $R_b >> R_{HS}$. Then the current will preferentially flow in the substrate as its distribution is governed by the ratio $R_b/R_{HS}$. In order to experimentally test this conjecture, the barrier resistance was estimated for the case of 10- and 20- filament samples from the data plotted in Fig. 8. The barrier resistance corresponds to the two steep "jumps" at the edges of the tape, and its value is $R_b \sim$ 1-1.5 mΩ both for the 10- and 20-filament samples. On the other hand, as we have seen above, the central part of the tape has a gradual linear increase of the resistance, whose slope is not far from the slope of Hastelloy only (continuous blue line in Figure 9). We therefore think that the current does not flow in and out of each filament, but flows prevalently in the Hastelloy.

We have seen in section 3 that oxidation has a positive effect in reducing the coupling losses. This positive effect is also observed in terms of change of transverse resistance. After oxidation we measured a transverse resistance between the first and the last filament of 18 mΩ, both for the 10- and 20-filament samples, which is 5 times higher than the value before oxidation (3.5 mΩ, see data points at the extreme right in Figure 9).

## 5. Conclusion

With this work we combined systematic sample preparation and a selection of structured samples to demonstrate that it is technologically possible to produce striated coated conductors with a large number of high-quality filaments (up to 120) without an excessive reduction of critical current: the samples with 80 and 120 filaments retain almost 80 and 70% of their current-carrying potential, respectively.

The striations are effective for reducing the hysteretic AC losses, but it is also clear that as the number of filaments increases the coupling loss becomes more important. As a consequence the total loss reduction is not proportional to the number of filaments as in the ideal case of uncoupled filaments: for example the 10-filament tapes has 10 times lower losses than the original tape (as expected for perfectly uncoupled filaments), but in the case of the 120-filament tape the loss reduction is by a factor 40.

We also investigated the transverse resistance between filaments, in order to get an insight on the way the coupling currents flow inside the conductors. The coupling currents path is governed by ratio between the barrier/interface ($R_b$) and the substrate ($R_{HS}$) resistance. We found that $R_b$ is 1-2 orders of magnitude higher than $R_{HS}$. Therefore the coupling current prevalently flows in the metallic substrate, rather than in and out of the filaments, as usually foreseen. By oxidizing the samples, the barrier/interface resistance between filaments and substrate was significantly increased and the filament coupling correspondingly reduced. In this case the loss reduction is much more substantial. For example, in the case of the 120-filament sample, it reaches a factor of 3 at 130 Hz.

The extension of this work will be to investigate technologically viable solutions to apply the same methods and analyses to samples with copper stabilization.

## Acknowledgments

This work was partly supported by the Helmholtz-University Young Investigator Group Grant VH-NG- 617.